\documentclass{iauc}
\usepackage{graphicx}

\title[Remote Star Clusters in NGC 6822] %% give here short title %% 
{Discovery of Remote Star Clusters in the Halo of the Irregular Galaxy NGC 6822}

\author[Hwang et al.]   %% give here short author list %%
{Narae Hwang$^1$, Myung Gyoon Lee$^1$, Jong Chul Lee$^1$, Won-Kee Park$^1$,
Hong Soo Park$^1$, Jang-Hyun Park$^2$, Sangmo Tony Sohn$^2$, \break Sang-Gak Lee$^1$,
Hyung Mok Lee$^1$, Mun-Suk Chun$^3$, \break Young-Wook Lee$^3$, Young-Jong Sohn$^3$,
In-Soo Yuk$^{1,2}$, \break Sang Chul Kim$^{1,2}$, Ho-Il Kim$^2$, and Wonyong Han$^2$}
\affiliation{$^1$Astronomy Program, SEES,
Seoul National University, Seoul 151-742, Korea \break
email: nhwang@astro.snu.ac.kr \\[\affilskip]
$^2$Korea Astronomy \& Space Science Institute, 61-1 Whaam Yuseong, Daejeon 305-348, Korea \break
$^3$Center for Space Astrophysics, Yonsei University, Seoul 120-179, Korea}

\pubyear{2005}
\volume{198} %% insert here IAU Symposium No.
\pagerange{119--126}
\date{?? and in revised form ??}
\jname{Near Field Cosmology with Dwarf Elliptical Galaxies}
\editors{H. Jerjen and B. Bingelli, eds.}

\begin{document}

\maketitle

\begin{abstract}
We report the discovery of three new star clusters in the halo of the Local Group
dwarf irregular galaxy NGC 6822. These clusters were found in the deep images
taken with the MegaPrime at the CFHT, covering a total field of 2 deg $\times$ 2 deg.
The most remote cluster is found to be located as far as 79 arcmin away
from the center of NGC 6822.
This distance is several times larger than the size of the
region in NGC 6822 where star clusters were previously found.
Morphological structures of the clusters and 
color-magnitude diagrams of the resolved stars in the
clusters show that at least two of these clusters are proabably old globular clusters.
%% add here a maximum of 10 keywords, to be taken form the file <Keywords.txt>.

\end{abstract}

\firstsection % if your document starts with a section,
\section{Introduction}

Star clusters in the dwarf irregular galaxy NGC 6822 %%($(m-M)_{0} \sim 23.48 \pm 0.08$) 
have been first reported by \cite{hubble25} and
systematically investigated by \cite{hodge77}.
The most recent study by \cite{KH04} used the available 
HST Archive data and cataloged all the star clusters identified
including one genuine globular cluster Hubble VII.
However, no star cluster survey in the outer halo of NGC 6822 has ever been tried before,
although the existence of a rather large stellar halo around the galaxy was 
suggested by \cite{letarte02} from their Carbon star survey.

\section{Discovery}
\begin{figure}
\includegraphics[scale=.30]{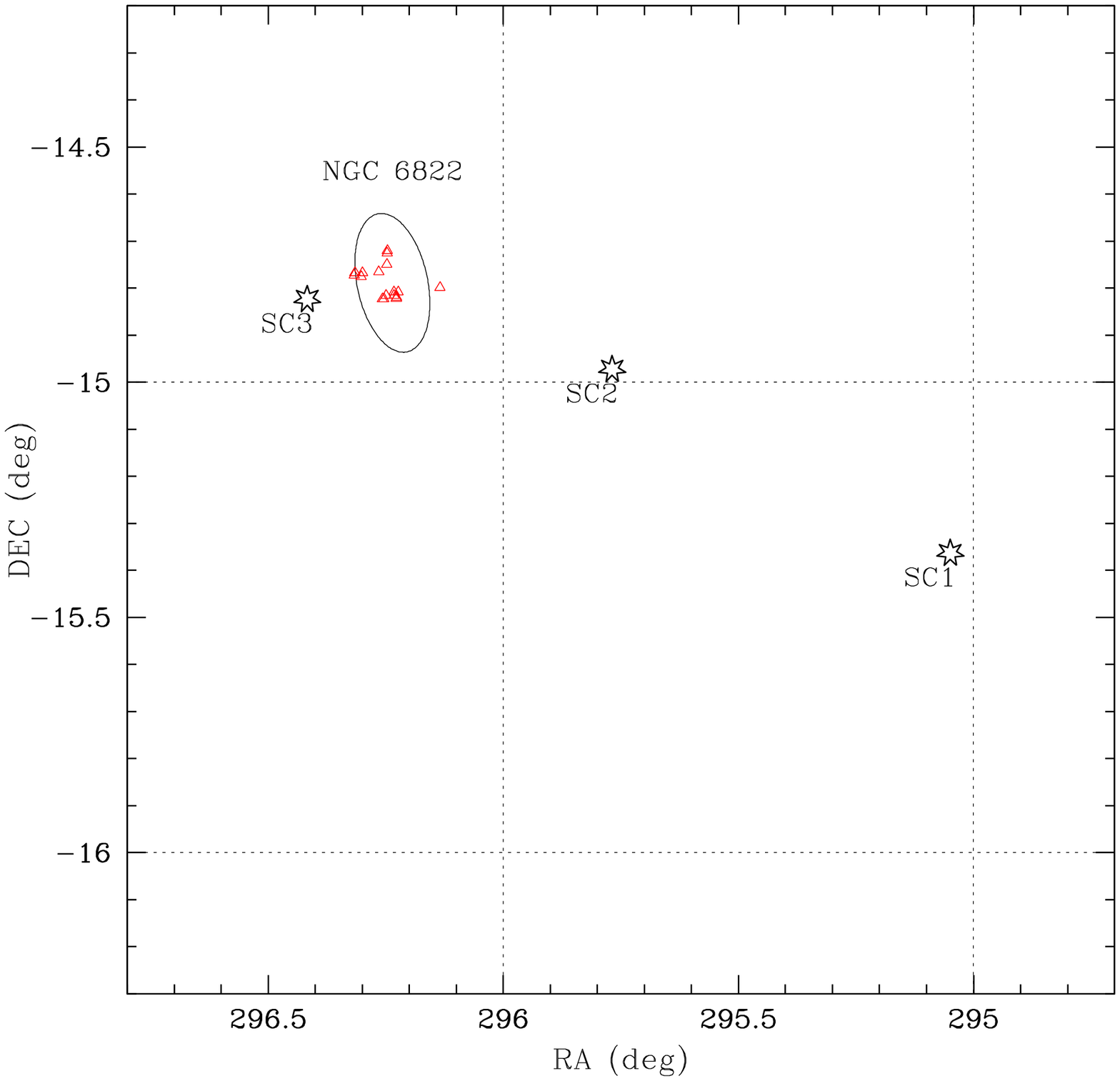}
\includegraphics[scale=.31]{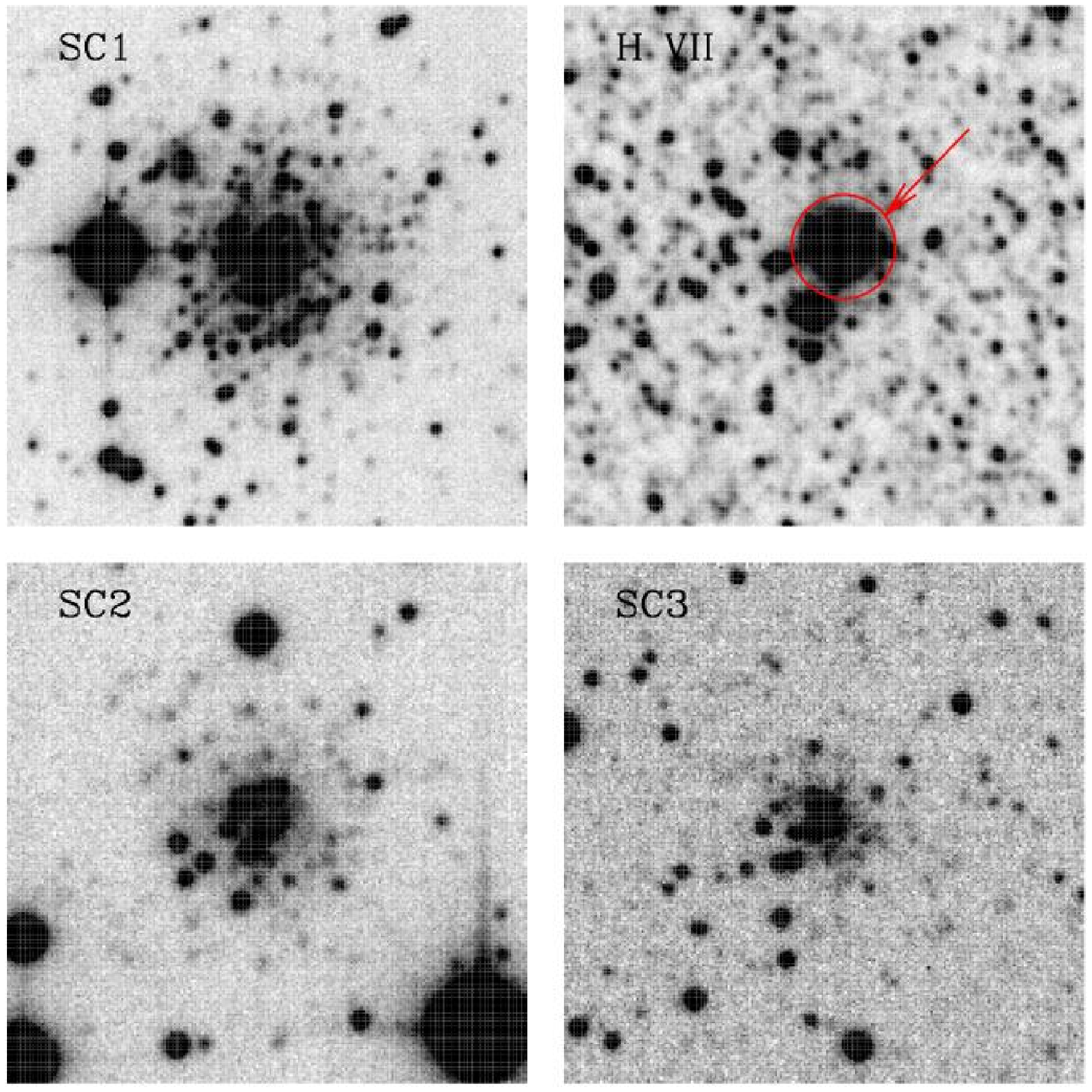}
  \caption{\textit{Left}: Locations of the new star clusters in 
NGC 6822 (stellar symbols). The SC1 is about 12 kpc
away from the galaxy center. Marks (triangles) inside and around the ellipse are previously 
known star clusters. 
\textit{Right}: Sloan \textit{i} band images of the three new 
star clusters and a known globular cluster Hubble VII in NGC 6822 
(upper right; marked by an arrow). Note the resolved member stars of 
new star clusters. Each image is $37'' \times 37''$ wide.}
  \label{fig}
\end{figure}

Visual inspection of %and detected source clustering effect investigation on 
the wide field survey data around NGC 6822 
has revealed three new star clusters.
The locations and morphologies of these clusters are shown in Figure~\ref{fig}.
From morphological and photometric studies,
two clusters, SC1 and SC2, are regarded genuine old globular clusters 
with ages more than 3 Gyrs (see \cite[Hwang et al. 2005]{hwang05} for details).
One noteworthy point is that the new star clusters are distributed to the 
very remote places (\textit{Left} panel of Figure~\ref{fig}).
The projected distance from the NGC 6822 center to SC1, 
the most remote cluster, is about 12 kpc. 
For comparison, NGC 1841, the outermost star cluster in the LMC, 
is located at about 13 kpc from the LMC center. 
Another important point in the images is that all three clusters are extended 
and are clearly resolved into stars, 
whereas Hubble VII is not resolved at all in our data
(\textit{Right} panel of Figure~\ref{fig}).
Further investigation shows that the half-light radii of these new clusters 
are larger than 10 pc and even larger than 20 pc for SC1.

\section{Implications}
The existence of the newly discovered star clusters suggests that the 
underlying halo has
a different structure from a giant HI disk-like cloud 
which is extended along NW-SE direction (\cite[de Blok \& Walter 2000]{BW2000}).
These clusters also provide a proof that the halo of NGC 6822 is quite larger than 
previously expected (see \cite[Lee \& Hwang 2005]{mglee05} in this volume).

The extended structures of these new clusters are very unusual features.
The SC1, among these, is found to be as extended as new 
star clusters recently discovered in the halo of M31 
(\cite[Huxor et al. 2005]{hux05}; \cite[Lee et al. 2005]{mglee05b}). 
The formation mechanism of these extended star clusters, 
including the correlation with evolutionary history of the host galaxies,
is not clearly understood yet, requiring further studies.

\begin{acknowledgments}
N.Hwang was supported in part by the BK21 program.
M.G.Lee was supported in part by the ABRL(R14-2002-058-010000-0).
\end{acknowledgments}

\end{document}